# Prevalence and incidence of postpartum depression and environmental factors: the IGEDEPP cohort

Sarah Tebeka, M.D[1,2]; Yann Le Strat, M.D, Ph.D[1,2]; Alix De Premorel Higgons, M.D[1,2]; Alexandra Benachi, M.D, Ph.D[3,4]; Marc Dommergues, M.D, Ph.D[5,6]; Gilles Kayem, M.D, Ph.D[5,7]; Jacques Lepercq, M.D[8,9]; Dominique Luton, M.D, Ph.D[8,10]; Laurent Mandelbrot, M.D, Ph.D[8,11]; Yves Ville, M.D, Ph.D[5,12]; Nicolas Ramoz, Ph.D[1]; Sophie Tezenas du Montcel, M.D, Ph.D[5,13]; IGEDEPP Groups*; Jimmy Mullaert, M.D, Ph.D[14,15]; Caroline Dubertret, M.D, Ph.D[1,2,]

[1] Université de Paris, INSERM UMR1266, Institute of Psychiatry and Neurosciences, Team 1, Paris, France

[2] Department of Psychiatry, AP-HP, Louis Mourier Hospital, F-92700 Colombes, France

[3] Université Paris Sud, Clamart, France

[4] Department of Obstetrics and Gynecology, AP-HP, Antoine Beclere Hospital, Clamart, France

[5] Sorbonne University, Paris, France

[6] Department of Obstetrics and Gynecology, AP-HP, Hôpital Universitaire Pitié-Salpêtrière, Paris, France

[7] Department of Obstetrics and Gynecology, AP-HP, Trousseau Hospital, Paris, France

[8] Université de Paris, Departement Hospitalier Universitaire Risks in Pregnancy, Paris, France

[9] Port-Royal Maternity Unit, Cochin Hospital, AP-HP, Paris, France.

[10] Department of Obstetrics and Gynecology, AP-HP, Bichat Hospital, Paris, France;

[11] Department of Obstetrics and Gynecology, AP-HP, Louis Mourier Hospital, Colombes, France;

[12] Department of Obstetrics and Gynecology, AP-HP, Necker enfant malade hospital, Paris, France

[13] INSERM, Institut Pierre Louis d'Epidémiologie et de Santé Publique IPLESP, AP-HP, Hôpitaux Universitaires Pitié Salpêtrière - Charles Foix, F75013 Paris, France

[14] Department of Epidemiology, Biostatistics and Clinical Research, AP-HP, Hôpital Bichat, F-75018 Paris, France

[15] Université de Paris, IAME, INSERM, F-75018 Paris, France

* The IGEDEPP group is made up of all the clinicians involved in the study :

Alexandra Benachi, Emmanuelle Bertin, Cecile Bourneuf, Jeanne Colombe, Laura Couppa, Marc Dommergue, Caroline Dubertret, Fanny Georges, Celine Hebbache, Gilles Kayem, Marie Le Bars, Yann Le Strat, Jacques Lepercq, Dominique Luton, Julie Guiot Madhavi, Laurent Mandelbrot, Jimmy Mullaert, Cindy Parent, and Alix de Premorel, Nicolas Ramoz, Sarah Tebeka, and Yves Ville.

**Corresponding Author:** Dr. Sarah Tebeka,

**ABSTRACT**

**Background:** IGEDEPP (Interaction of Gene and Environment of Depression during PostPartum) is a prospective multicenter cohort study of 3,310 Caucasian women who gave birth between 2011 and 2016, with follow-up until one year postpartum. The aim of the current study is to describe the cohort and estimate the prevalence and cumulative incidence of early and late postpartum depression (PPD).

**Methods**: Socio-demographic data, personal and family psychiatric history, as well as stressful life events during childhood and pregnancy were evaluated at baseline. Early and late PPD were assessed at 8 weeks and 1 year postpartum respectively, using DSM-5 criteria.

**Results**: The prevalence of early PPD was 8.3% (95%CI 7.3-9.3), and late PPD 12.9% (95%CI 11.5-14.2), resulting in an 8-week cumulative incidence of 8.5% (95%CI 7.4-9.6) and a one-year cumulative incidence of PPD of 18.1% (95%CI: 17.1-19.2). Nearly half of the cohort (N=1571, 47.5%) had a history of at least one psychiatric or addictive disorder, primarily depressive disorder (35%). Almost 300 women in the cohort (9.0%) reported childhood trauma. During pregnancy, 47.7% women experienced a stressful event, 30.2% in the first 8 weeks and 43.9% between 8 weeks and one year postpartum. Nearly one in five women reported at least one stressful postpartum event at 8 weeks.

**Conclusion**: Incident depressive episodes affected nearly one in five women during the first year postpartum. Most women had stressful perinatal events. Further IGEDEPP studies will aim to disentangle the impact of childhood and pregnancy-related stressful events on postpartum mental disorders.

**HIGHLIGHTS**

- The prevalence of early PPD was 8.3% (95%CI 7.3-9.3) and the prevalence of late PPD was 12.9% (95%CI 11.5-14.2)
- The eight-week cumulative incidence of PPD was 8.5% (95%CI 7.4-9.6) and one-year cumulative incidence of PPD was 18.1% (95%CI: 17.1-19.2).
- Nearly half of the 3310 women (N=1571, 47.5%) had a history of at least one psychiatric disorder.

- Of the 3310 women, 298 (9.0%) reported childhood trauma and most of them experienced stressful perinatal events.
- During pregnancy, 47.7% women experienced a stressful event with a high negative impact and 43.9% between 8 weeks and one year postpartum.

**Funding:**

The study was funded by a grant from the Programme Hospitalier de Recherche Clinique - PHRC 2010 (French Ministry of Health). The study was sponsored by Assistance Publique – Hôpitaux de Paris (Délégation à la Recherche Clinique et à l'Innovation).

**Authors' contributions:**

S. Tebeka drafted the initial manuscript, and approved the final manuscript as submitted.

J. Mullaert carried out the initial analyses, revised the manuscript, and approved the final manuscript as submitted.

C. Dubertret designed the study, revised the manuscript, and approved the final manuscript as submitted.

Y. Le Strat, A. De Premorel Higgons, A. Benachi, M. Dommergues, G Kayem, J. Lepercq, D. Luton, L. Mandelbrot, Y. Ville, N. Ramoz, and S. Tezenas du Montcel revised the manuscript, and approved the final manuscript as submitted.

**Competing interests:** none of the authors have any competing interests.

**Acknowledgments**:

We thank all the clinicians involved in this study, especially Cindy Parent, Julie Guillon, Jeanne Colombe, Cecile Bourneuf, Celine Hebbache, Madhavi-Julie Guiot, Laura Couppa and Marie Lebars who recruited and followed the participants

We thank all the women who participated in the study.

**INTRODUCTION**

Pregnancy and postpartum comprise a timeframe in which the prevalence and the burden of depression is high (Howard et al., 2014; Mann et al., 2010; Tebeka et al., 2016). Postpartum depression (PPD) is a heterogeneous disease, both in terms of its symptoms as well as their onset (PACT Consortium, 2015; Putnam et al., 2017). Postpartum depression (PPD) is defined by international diagnostic criteria as a major depressive episode starting before the four (for DSM-5) or six week mark (for ICD-11) after delivery (American Psychiatric Association, 2013; World Health Organization, 2018). However, a broader definition, with a range up to one year postpartum is commonly used. This definition is also retained by the international Marce society of perinatal mental health (The International Marce Society, 2015). Although the definition remains controversial, some authors have proposed a definition of early PPD as beginning before the end of the first two months following birth, and late PPD as depression beginning between two months to one year after birth. (Gjerdingen et al., 2011; Norhayati et al., 2015).

PPD can have deleterious consequences for the mother. Suicide is the cause of more than 4% of maternal deaths (Metz et al., 2016). Forty percent of women who have experienced PPD will have a recurrence of depression within their lifetime (Howard et al., 2014); nearly 50% will experience another episode of PPD in subsequent pregnancies (Goodman, 2004). PPD is also associated with adverse outcomes for children (Netsi et al., 2018), including poor quality of early interactions with their mothers, pervasive effects upon their cognitive, emotional, and social development (Grace et al., 2003), and an increased risk for psychiatric disorders during infancy, childhood, adolescence and adulthood (Murray et al., 2011). PPD is therefore a major public health issue (Wisner et al., 2006).

PPD is considered to be a multifactorial disease, combining both environmental risk factors (including stressful life events), and genetic risk factors for depression (Howard et al., 2014). A better understanding of the pathophysiological mechanisms of PPD may help in targeting women at risk of PPD and developing programs in prevention and intervention. During the last decade, a wide range of research has helped to disentangle a number of factors associated with PPD, including pregnancy-related factors (gestational diabetes (Mak et al., 2018), hypertensive disorders (e.g. systolic blood pressure ≥140 mm Hg or diastolic blood pressure ≥90 mm Hg, or proteinuria) (Strapasson et al., 2018) and stressful maternal experiences (such as sexual assault, physical violence, and emotional distress) (Zhang et al., 2018).

There are different methods to diagnose PPD: structured interviews, such as the Composite International Diagnostic Interview (Robins et al., 1988) or the Mini International Neuropsychiatric Interview (Lecrubier et al., 1997), semi-structured interviews such as the Structured Clinical Interview for DSM disorders (First, 2015), or Diagnostic Interview for Genetic Studies (DIGS) (Nurnberger et al., 1994), or self-questionnaires such as Edinburgh Postnatal Depression Scale (EPDS) (Chaudron et al., 2010; Cox et al., 1987). Structured interviews administered by interviewers without any training; without a clinician's assessment, may induce an overestimation of depression (Levis et al., 2019, 2018). Self-questionnaires are easy to use and well accepted. EPDS is the most validated and widely used tool for peripartum depression in the world(Hewitt et al., 2010). However, it is a self-questionnaire validated only for PPD screening and includes different validated cutoff scores depending on whether it's being used in studies, across different countries, or in the ante or postnatal period of assessment (Hewitt et al., 2010; Norhayati et al., 2015; Smith-Nielsen et al., 2018). Although semi-structured interviews are time-consuming and require clinically trained interviewers, we favored this approach which allows for a reliable diagnosis, taking into account clinical judgment for each symptom of PPD.

Based on a large prospective multicentric cohort of women in the postpartum period, our aims were threefold: (1) to describe the IGEDEPP cohort (Interaction of Gene and Environment of Depression during PostPartum) developed to identify risk factors for PPD, (2) to estimate the cumulative incidence and prevalence of PPD at 8 weeks and one year after delivery, and (3) to estimate the prevalence of lifetime psychiatric disorders (mood, anxiety, eating and substance use disorders and suicide attempts) and the frequency of stressful events during childhood, pregnancy, and postpartum.

**METHODS**

### Participants

IGEDEPP is a prospective cohort of 3,310 Caucasian women who gave birth in eight maternity departments in the Paris metropolitan area in France, between November 2011 and June 2016. The women included in this cohort were aged over 18, Caucasian, French speaking, and were covered by French social insurance. The sample was restricted to Caucasian women to ensure ethnic homogeneity for genetic studies. Women were defined as Caucasian if they self-declared that their four grandparentswere Caucasian. Exclusion criteria were delivery before 32 weeks of gestation, schizophrenia or mental retardation which were assessed clinically according to international criteria. Women meeting the inclusion criteria were visited by a clinician who presented the study at the maternity department between the second and the fifth day after delivery. After the patient provided consent, she had a blood test and a 30- to 90-minute interview with the clinician. No information was obtained from the medical record; all information was collected during the interview with the clinician. Acceptance rate was 61.2%. No data was collected on any woman who refused participation.

The research protocol (ClinicalTrials.gov Identifier: NCT01648816), including informed consent procedures, was approved by the French Ethics committee (Ile de France I) and Data Protection and Freedom of Information Commissions.

**Measures**

Women were evaluated by a clinician at three points over the course of one year. The interviewers were psychiatrists or psychologists who received specific training to administer the semi-structured interview of the DIGS, and to assess patients' psychiatric history, as well as their life events. They benefited from training on all scales of the study. The first face-to-face interview took place at the maternity department between the second and fifth day after delivery. The second and third interviews were conducted via phone calls at 8 weeks and one year postpartum. All assessments are summarized in table 1.

**Screening tools and diagnosis of early and late PPD**

All women were assessed for PPD in concordance with DSM-5 criteria for major depressive disorder(American Psychiatric Association, 2013). Depression was evaluated at three time points, using the DIGS: at inclusion for the diagnosis of current depression and a history of depression, at 8 weeks postpartum for the diagnosis of early PPD, and at one year postpartum for late PPD. Late PPD was defined by the presence of a major depressive episode with onset between two months and one year postpartum. Women were assessed for the onset of depressive symptoms in the past 10 months and their duration. Thus, late PPD was assessed retrospectively one year after delivery, and could persist at the time of assessment or not.

All women also were screened for PPD at each of the three time points by the Edinburgh Postpartum Depression Scale (EPDS) (Cox et al., 1987; Guedeney and Fermanian, 1998), the most widely used self-administered questionnaire for PPD, and by two scales used in the general population for depression: the Hospital Anxiety and Depression Scale (HAD) (Zigmond and Snaith, 1983), a hetero questionnaire, and the Center for Epidemiologic Studies-Depression (CES-d) scale (Radloff, 1977), a self-administered instrument. Although all of the women in our sample had completed the EPDS at each of the three evaluations, the diagnosis of PPD according to DSM-5 criteria for major depressive disorder is based solely on the clinical evaluation of the DIGS interview.

To complete the mood assessment, we used the French version of the "maternity blues" questionnaire, a 28-item self-administered questionnaire to assess the intensity of postpartum blues on days 2 to 5 postpartum (Glangeaud-Freudenthal et al., 1995).

**Sociodemographic Measures at the first evaluation**

Age, marital status, educational level, employment, and health insurance status were systematically collected. Age at interview was categorized into: (i) 18–25, (ii) 26-34, (iii) 35-39, and (iv) 40 years or more. Marital status was assessed using six categories, grouped into two clusters: (i) widowed, divorced, separated, never married and (ii) married and common-law married. Educational level was classified into: (i) high school level or below, (ii) university level or higher. Employment status was categorized into: (i) employed and (ii) unemployed. Health insurance status was categorized into (i) covered under social security, the usual health insurance scheme and (ii) covered under universal healthcare system (PUMA) free healthcare available to low income families.

**Lifetime psychiatric disorders assessed at the first evaluation**

All women were evaluated by a trained clinician (psychologist or psychiatrist) using the Diagnostic Interview for Genetic Studies (DIGS) (Nurnberger et al., 1994; Preisig et al., 1999), a semi-structured interview assessing depression and other psychiatric or substance use disorders according to DSM-IV-TR criteria over the lifetime. Regarding the diagnosis of depression, we applied DSM-5 criteria for all depressive disorders including depression with mixed characteristics (which were precisely sought using of the DIGS), but excluding bereavement as a criterion (American Psychiatric Association, 2013). All clinicians were specifically trained to administer the DIGS interview. We reported lifetime prevalence for mood disorders (including major depressive episode, and bipolar disorder), anxiety disorders (including panic disorder, agoraphobia, social anxiety disorder, specific phobia, social phobia, generalized anxiety disorder and obsessive-compulsive disorder), substance use disorders (including those related to tobacco, alcohol, cannabis, cocaine, stimulants, opiates, hallucinogens and other drugs), as well as eating disorders (anorexia and bulimia). Regarding substance use, a pregnancy-specific assessment was proposed to assess tobacco, alcohol and cannabis use specifically in the first, second and third trimesters of pregnancy. Lifetime suicide attempts were also assessed.

First-degree family history (parents and siblings) of psychiatric disorders was obtained using the Family Informant Schedule and Criteria (FISC) (Andreasen et al., 1977) assessing mood disorders, anxiety disorders, schizophrenia, alcohol and substance use disorders for all first degree relatives. Participants reporting at least one first degree relative with a psychiatric disorder were considered as having a family history of the given disorder.

**Stressful life events during childhood, pregnancy, and postpartum.**

Childhood trauma was investigated using a self-administered questionnaire, the Childhood Trauma Questionnaire (CTQ), composed of 28 questions encompassing 5 different types of childhood trauma: sexual abuse, emotional or physical neglect and abuse (Bernstein et al., 2003). Cut-off scores were defined by Paquette et al. (Paquette et al., 2004) to determine the presence of abuse/neglect: > 10 for physical abuse and sexual abuse sub-scores, > 13 for physical neglect sub-scores, and > 15 for emotional abuse and emotional neglect sub-scores.

To identify stressful life events occurring during pregnancy and postpartum, we used the Paykel scale (Paykel, 1997), a 64 item questionnaire measuring the subjective impact of each event along with the extent of life change related to the event. These items are organized into categories (work, education, wealth, health, bereavement, relocation, romantic relationships, legal issues, family, marital and other). Stressful life events were defined as events rated by participants as having a "marked" to "severe" negative impact (impact score of 1 to 2). In other words, stressful events were defined by the subjective experience of the event; and not the objective severity of the life event itself. Paykel scales were administered at 3 time points: at the maternity department for the evaluation of stressful life events during pregnancy, by phone at 8 weeks and 1 year postpartum for the evaluation of stressful events during weeks 0-8 and 8-52 respectively. For each category, we reported the number and proportion of participants who declared at least one stressful event.

To complete the assessment of stressful life events, the BRIEF COPE was used to evaluate inappropriate coping strategies for stressful situations (Carver, 1997; Muller and Spitz, 2003).

**Obstetrical characteristics during pregnancy and postpartum**

Obstetric characteristics were collected at the maternity department by a hetero-questionnaire evaluating the following subjects: past history of pregnancy, use of assisted reproductive technology (ART), existence of multiple pregnancy, existence of a chronic underlying disease, medication or emergency consultation or hospitalization during pregnancy, and the reason for hospitalization: hypertension, threatened preterm delivery, gestational diabetes, thromboembolic event. Data on delivery included: labor onset (spontaneous or induced), mode of delivery (vaginal or cesarean (before or during labor)), use of obstetrical analgesia and perineal trauma (none, episiotomy, or perineal trauma if ≥ 2-degree perineal tear).Prematurity (< 37 weeks of gestation), low birth weight (<2500g), presence of postpartum hemorrhage, use of the intensive care unit for the baby or the mother, and information on breastfeeding was collected.

At the 8 weeks postpartum assessment, information on the following postpartum events was collected: hospitalization for baby or mother, pain, incontinence, sexual activity and breastfeeding. At one year, we assessed postpartum events including hospitalization for baby or mother, breastfeeding and new pregnancy.

We asked the women about the impact of all obstetrical events: stressful obstetrical events are those with a "marked to severe" negative impact.

*Statistical Analyses*

The variables describing the included population were reported, for binary variables, as counts and percentages. For quantitative variables, descriptive statistics were mean and standard deviations or median and interquartile range. Cumulated incidence estimates were reported along with their 95% confidence intervals (95 % CI) derived by Kaplan-Meyer survival analysis. Follow-up time for cumulative incidence started at delivery and ended at the date of last contact (for participants without depression) and at the reported beginning of symptoms for participants with depression. Comparisons of baseline characteristics between participants lost to follow-up and others were performed by logistic regression. Odd Ratios (ORs) and their 95 % CI are reported with the asymptotic p-value from Wald's statistic.

All computations were done with the software R, version 3.4.

**RESULTS**

At the maternity departments, 3310 women were included. Of these 3015 (91.1%) were assessed at 8 weeks postpartum and 2351 (71.0 %) women were followed-up at 1 year postpartum (**Figure 1**).

Participants lost to follow-up were significantly younger (p<0.05) and less educated (p<0.05) than those who were not lost to follow-up. Other baseline characteristics were not significantly different among the two groups, including lifetime history of major depressive episode (**Supplementary tables 1** and **2**).

**PPD prevalence and cumulative incidence (figure 2)**

Among the 3015 women assessed at 8 weeks postpartum, 250 (8.3%, 95%CI 7.3-9.3) met criteria for early PPD. Among the 2351 women assessed at one year postpartum, 304 (12.9 %, 95%CI 11.6-14.3) met criteria for late PPD, 68 of whom already had early PPD. Before 8 weeks postpartum, 256 incident cases of PPD were recorded, resulting in a cumulative incidence of 8.5% (95%CI 7.4-9.6). The prevalence was heterogeneous among centers (**Supplementary table 3).** A total of 481 women presented with PPD during the first year after delivery and the one-year cumulative incidence of PPD was 18.1% (95%CI: 17.1-19.2). **Figure 2** presents the incidence curve over time during the first year. Among the women who completed the three assessments, 1875 were not diagnosed with a major depressive episode in the period of pregnancy and postpartum.

### Sociodemographic characteristics at the first evaluation (table 2)

The majority of women in our sample were between 26 and 34 years old (62.7 %), with an average age of 32 years. Most of the participants were currently married or in domestic partnerships (common-law married) (96.7 %), employed (93.3 %), and had a high level of education (92.0 %).

### Family and personal lifetime psychiatric history at the first evaluation (tables 3 and 4)

Nearly half of the women reported at least one lifetime psychiatric or addictive disorder (n=1571, 47.5%). More than one third of women had a history of mood disorder (n=1168, 35.3 %), mainly major depressive episode. Only 6 (0.2%) of our sample had a history of bipolar disorder. Major depressive episode during the current pregnancy concerned 3.7 % of the whole sample. Ninety-eight women (3.0 %) reported at least one suicide attempt. A personal history of anxiety disorder was reported by 546 (16.0 %) participants, including specific phobia (7.2 %), agoraphobia (5.1 %), social anxiety (4.2 %), generalized anxiety disorder (2.1 %), panic disorder (1.5 %) and obsessional compulsive disorder (0.2 %). A lifetime history of eating disorder was reported by 4.1 % of participants, including anorexia (3.7 %) and bulimia (1.2 %).

Substance use disorder affected 8.5 % of women, mainly tobacco use disorder (7.1 %). Excluding tobacco, substance use disorders were reported by 2.3 % of the sample: 1.7 % of women had a lifetime history of cannabis use disorder, 0.6 % presented a lifetime history of alcohol use disorder and 0.4 % reported other lifetime substances use disorders (cocaine, stimulant, opioids, hallucinogens and others). During pregnancy, 42% of women (n=1387) drank at least once alcohol and 0.32% met criteria for an alcohol use disorder. In our cohort, 16% of women had smoked tobacco during pregnancy, while 0.85% of women reported having used cannabis, and 0.21% had a cannabis use disorder during pregnancy.

The majority (n=2141, 64.7 %) of women had a family history of psychiatric disorder with mood disorders being the most frequent (reported by 49.0 % of women), followed by family history of anxiety disorder (21.0 %), family history of alcohol use disorder (15.1 %) and family history of schizophrenia (1.4 %).

**Stressful life events during childhood, pregnancy, and postpartum period (tables 5 and 6)**

About one in ten woman (9.0 %) reported childhood trauma. The most frequent kind of trauma reported was emotional neglect (5.9 %), followed by sexual abuse (2.6 %), physical abuse (2.1 %), emotional abuse (2.0 %) and physical neglect (0.79 %) (**Table 5**).

During pregnancy, 85.0 % of women experienced at least one event listed on the Paykel scale. A total of 1578 (47.7 %) women experienced a stressful event with a high negative impact (**Supplementary table 4**). During the 8-week follow-up, 912 (30.2 %) women experienced at least one stressful event with a negative impact. During the 1-year follow-up, 1033 (43.9 %) reported at least one stressful with a negative impact.

**Obstetrical characteristics during pregnancy and postpartum (table 6)**

Assisted reproductive technology was necessary for 269 (8.1 %) women. Among our sample, 1242 (37.5 %) women were primiparous, 105 (3.2 %) had a multiple pregnancy and 439 (13.3 %) women had been hospitalized during pregnancy. In the IGEDEPP sample, 1982 (59.9 %) women had spontaneous labor onset; 2493 women (24.7 %) had a cesarean delivery. Among the women, 516 (15.6 %) intended to deliver with obstetrical analgesia but ultimately delivered without, 782 (23.6) had an episiotomy and 30 (0.9 %) had perineal trauma. Delivery events with a negative impact concerned 846 (25.6 %) of women. Those events included preterm birth (4.8 %), low birth weight (6.3 %), intensive care for the baby or the mother (1.1 % each), and postpartum hemorrhage (4.6 %).

At 8 weeks, 637 (21.1 %) women had at least one postpartum event with a negative impact. The main events reported were hospitalization for the child (11.0 %) or the mother (1.3 %), pain or incontinence (25.7 %) and the absence of sexual activity (30.0 %). One year after birth, 419 (17.8 %) women had at least one postpartum event with a negative impact. The main events reported were hospitalization or disease of the child (15.9 %), or the mother (1.4 %). A new pregnancy concerned 4.1 % of women.

Breastfeeding after birth concerned 2372 women (71.7 %) and the median duration of breastfeeding was 12 weeks (interquartile range 7-24). The prevalence of breastfeeding was 52.8 % at 8 weeks and 8.2 % at one year postpartum.

**DISCUSSION**

This study aimed to describe IGEDEPP data for the first time. IGEDEPP is the largest prospective multicenter cohort to our knowledge of 3310 pregnant women followed-up for 1 year to determine the prevalence of early and late postpartum depression (PPD) during the first year after delivery. In the first postpartum year, the cumulative incidence of PPD was 18.1 %. Early PPD affected 250 women (8.3 %), and 304 women had late PPD (12.9 %). These results are consistent with different international studies, evaluating PPD prevalence between 10 and 20 % during the first year postpartum (Ko et al., 2012; Heron et al., 2004; Shorey et al., 2018). Many factors can explain this range: the studies are heterogeneous in terms of population, timing of assessment and diagnostic tools (Norhayati et al., 2015). For instance, we used a diagnosis based on clinical evaluation with DSM-5 criteria, whereas most studies rely only on a screening questionnaire, leading to possible fluctuations in PPD prevalence.

Our sample has very favorable socio-demographic characteristics: the women included had a high level of education; and the vast majority of them worked and was in a relationship. Compared with the ALSPAC study, an English cohort, the women in IGEDEPP were more likely to be in a relationship (96.7% vs

79.4%) (Fraser et al., 2013). Compared with the 2016 French National Perinatal Survey (Blondel et al., 2017; French Ministry of Solidarities and Health, 2017), a national sample of births, women included in IGEDEPP were older (mean age was 30.3 vs. 32 years old in our sample) and had a higher level of education (55.4% vs. 92%). Despite its favorable socio-economic characteristics, the incidence of PPD remains high in our sample. A recent literature review highlighted conflicting results for the association between social status and risk of PPD (Norhayati et al., 2015). In developed countries, some authors have shown an association between a low level of education or income and PPD (Eastwood et al., 2012; Kim et al., 2008; Kozinszky et al., 2011), while others did not find this association (Deng et al., 2014; Goker et al., 2012; Hamdan et al., 2012; Leigh and Milgrom, 2008).

Nearly half of the women reported at least one lifetime psychiatric disorder, and one third of the women had a history of mood disorder. Prior research from the EDEN mother-child French cohort) found a history of mental health problems for 11% at 12.6% of women of childbearing age (Melchior et al., 2012; van der Waerden et al., 2015). This prevalence was based on a single question about psychiatric history, which may explain why it is much lower than that of our cohort. Indeed, we found similar rates of personal psychiatric history to those described in the National Epidemiologic Study of Alcohol and Related Conditions (NESARC), a study of 1085 peripartum women representative of the United States population assessed with DSM criteria, (48 % in both samples). However, French women of the IGEDEPP study were more likely to report a personal history of mood depressive disorder (35 % vs. 22 %) and more familial history of mood disorder (49 % vs. 37 %) than US women (Tebeka et al., 2016). One of the explanations could be that the two population (US vs French) and the assessment were different (depression in the past year vs prospective assessment in the first year postpartum).

Moreover, the same percentage of women reported at least one stressor event during pregnancy (85 % vs. 82 %, respectively) in our sample and in NESARC study (Tebeka et al., 2016).

**Strengths**

Our study has several strengths. First, to our knowledge, it is the largest multicenter prospective study of postpartum women which aims to evaluate the development of depression at 3 time points during the first year after birth. Moreover, depression diagnosis was established by a trained clinician using reliable DSM-5 criteria and does not rely solely on the use of a screening questionnaire, such as EPDS.

Furthermore, IGEDEPP provides an extensive assessment of a wide range of psychiatric disorders, including mood and anxiety disorders, substance use disorders and many stressful life events during childhood, pregnancy and one-year postpartum, as well as a broad assessment of obstetric events. Very few participants were lost to follow-up during the study: we had 91% available data at 8 weeks postpartum evaluation, and 71% at 1 year, increasing the external validity of these findings. However, because we have no data on those who refuse to participate to this cohort, we cannot avoid an inclusion bias of subjects with PPD that could change the incidence and prevalence.

**Limitations**

Some limitations should be considered when interpreting these results. First, participants in the IGEDEPP study reported higher levels of education than in the general population (INSEE, 2018), a finding which has also been associated with genetic research participation (Hensley Alford et al., 2011; McQuillan et al., 2003). This limits the generalizability of our findings. The eight inclusion centers were in Paris or in the Paris suburbs. This population is not representative of France. A general population study comparing a population from Ile de France to one from low Normandy found differing rates of depression: severe depression during the last year was more common in urban areas, and severe lifetime depression was more frequent in individuals living in flats in large building complexes. However, after adjusting for sex, age, marital status and negative life events, there was no longer any difference in the prevalence of depression in the two populations (Kovess-Masfety et al., 2005). Further study should examine how to reach the population currently excluded from genetic studies. Second, the sample is composed only of Caucasian women, in line with the design of the IGEDEPP study which includes a genetic component. Our study did not include French women of non-Caucasian ascent or migrant women and as migration is a known risk factor of postpartum depression (Ahmed et al., 2008; Falah-Hassani et al., 2015), we may have underestimated the prevalence of PPD in the general French population and lost information on risk factors in this vulnerable sub-group. Third, we only included women aged 18 or more. Adolescent pregnancy is associated with an elevated risk of depression and other psychiatric disorders and is associated with specific social and environmental conditions. Further study should examine whether the evidence provided by the IGEDEPP cohort apply to adolescent pregnancy or not. In addition, we used the CTQ to evaluate childhood trauma. The CTQ has been validated in a population of women of childbearing age, regardless of their pregnancy or postpartum status (Grassi-Oliveira et al., 2014; Paquette et al., 2004). According to a recent literature review , the CTQ is the most relevant tool for evaluating childhood trauma in the perinatal period (Choi and Sikkema, 2016). Finally, different clinicians have participated in the evaluation of patients, and although they have all benefited from specific training, we cannot exclude bias.

**Perspectives**

Incident PPD affects 18.1 % of women in the first year postpartum. Thanks to the evaluation of many different factors, including socio-demographics, family and personal psychiatric history, stressful life events during childhood, negative pregnancy, postpartum and obstetric events, the IGEDEPP study will be able to better characterize the factors associated with early and late PPD. In addition, genetic analyses with both gene candidates and a GWAS approach aim to clarify the genetic determinants of PPD. Finally, analysis of the interaction between genetic variants and life events will allow to accurately describe factors associated with PPD. Identifying a vulnerable population is necessary to develop prevention and intervention strategies during the pregnancy and postpartum period.

**LIST OF ABBREVIATIONS**

95%CI: 95% confidence interval

ART: assisted reproductive technology

CTQ: Childhood Trauma Questionnaire

EPDS: the Edinburgh Postnatal Depression

IGEDEPP: Interaction of Gene and Environment of Depression during PostPartum

ORs: Odd Ratios

PPD: Postpartum Depression

|  | Baseline | At 8 weeks | At one year |
|---|---|---|---|
| **Socio demographic assessment** | X | | |
| **Diagnostic Interview for Genetic Studies** (DIGS) (Nurnberger et al., 1994) | X | X [a] | X [b] |
| **Family Informant Schedule and Criteria** (FISC) (Andreasen et al., 1977) | X | | |
| **Edinburgh Postpartum Depression Scale** (EPDS) (Cox et al., 1987; Guedeney and Fermanian, 1998) | X | X | X |
| **Hospital Anxiety and Depression Scale** (HAD) (Zigmond and Snaith, 1983) | X | X | X |
| **Maternity blues Questionnaire** (Kennerley and Gath, 1989) | X | | |
| **Childhood Trauma Questionnaire** (CTQ) (Bernstein et al., 2003; Paquette et al., 2004) | X | | |
| **Scaling of life events (Paykel scale)** (Paykel, 1997) | X | X | X |
| **Obstetrical events** | X | X | X |
| **Brief COPE** (Carver, 1997; Muller and Spitz, 2003) | X | | |
| **CES-d** (Radloff, 1977) | X | | |

**Table 1:** measures made at baseline, 2nd and 3rd interviews

[a] only depression assessment.

[b] only depression and mania assessment.

Abbreviations: CES-d, Center for Epidemiologic Studies Depression Scale; COPE, Coping Orientation to Problems Experienced;

|                                               | N (%)        |
|-----------------------------------------------|--------------|
| Age (y)[a]                                    |              |
|     25 or less            | 158 (4.8)    |
|     26-34                 | 2077 (62.7)  |
|     35-39                 | 826 (25.0)   |
|     40 or more            | 248 (7.5)    |
| Marital status                                |              |
|     Widowed, divorced, separated, never married | 108 (3.3) |
|     Married, common-law married | 3202 (96.7) |
| Education level[b]                            |              |
|     University            | 3044 (92.0)  |
|     Primary or high school | 261 (7.9)   |
| Employment[c]                                 |              |
|     Unemployed            | 222 (6.7)    |
|     Employed              | 3088 (93.3)  |
| Health insurance[b]                           |              |
|     Social security       | 3281 (99.1)  |
|     Universal healthcare system (PUMA) | 24 (0.7) |

**Table 2:** Sociodemographic characteristics among IGEDEPP cohort respondents (N=3310)
[a] Data missing for 1 participant; [b]Data missing for 5 participants; [c]Data missing for 1 participant;
Abbreviations: PUMA, Protection Maladie Universelle ;

|  |  | N (%) |
|---|---|---|
| Mood disorders (any) |  | 1168 (35.3) |
|  | Major depressive episode | 1166 (35.2) |
|  | Bipolar disorder | 6 (0.2) |
| Suicide attempt |  | 98 (3.0) |
| Anxiety disorders (any) |  | 546 (16.0) |
|  | Obsessive-compulsive disorder | 8 (0.2) |
|  | Panic disorder | 51 (1.5) |
|  | Agoraphobia | 169 (5.1) |
|  | Social anxiety disorder | 138 (4.2) |
|  | Specific phobia | 238 (7.2) |
|  | Generalized anxiety disorder | 70 (2.1) |
| Eating disorders (any) |  | 137 (4.1) |
|  | Anorexia | 123 (3.7) |
|  | Bulimia | 40 (1.2) |
| Substance use disorders (any) |  | 280 (8.5) |
|  | Alcohol use disorder | 20 (0.6) |
|  | Cannabis use disorder | 55 (1.7) |
|  | Tobacco use disorder | 235 (7.1) |
|  | Other substance use disorder | 12 (0.4) |

**Table 3:** Lifetime psychiatric history among IGEDEPP cohort respondents (N=3310)

|                                                  | N (%)       |
|--------------------------------------------------|-------------|
| Family history of mood disorder                  | 1624 (49.0) |
| Family history of anxious disorder               | 696 (21.0)  |
| Family history of schizophrenia                  | 47 (1.4)    |
| Family history of alcohol dependence or abuse    | 501 (15.1)  |
| Family history of other substance use disorder   | 647 (19.5)  |

**Table 4:** Family history of psychiatric disorder among IGEDEPP cohort respondents (N=3310)

|  | N (%) |
|---|---|
| Childhood stressful life event | 298 (9.0) |
|     Emotional abuse | 100 (3.0) |
|     Physical abuse | 71 (2.1) |
|     Sexual abuse | 86 (2.6) |
|     Emotional neglect | 196 (5.9) |
|     Physical neglect | 26 (0.8) |

**Table 5:** Childhood stressful life events among IGEDEPP cohort respondents (N=3310)

| **Events before and during pregnancy** (N=3310) | |
|---|---|
| Infertility | 401 (12.1) |
| Assisted reproductive technology | 269 (8.1) |
| Physical concomitant chronic disease | 105 (3.2) |
| Primiparity | 1242 (37.5) |
| Multiple pregnancy | 105 (3.2) |
| Emergency consultation during pregnancy | 1593 (48) |
| Hospitalization during pregnancy | 439 (13.3) |
|     Threatened preterm delivery | 139 (4.2) |
|     Hypertension during pregnancy | 37 (1.1) |
|     Gestational diabetes | 59 (1.8) |
|     Venous thromboembolic event | 7 (0.21) |
| **Delivery events** (N=3310) | |
| Spontaneous labor onset | 1982 (59.9) |
| C-section delivery | 817 (24.7) |
| Absence of an accompanying person | 91 (2.5) |
| No obstetrical analgesia despite intention | 516 (15.6) |
| Perineal Trauma | |
|     None or minor perineal tear | 1674 (50.6) |
|     Episiotomy | 782 (23.6) |
|     Perineal trauma (≥ 2-degree perineal tear) | 30 (0.9) |
| Newborn related events (preterm, low weight, NICU) | 310 (9.4) |
| Maternal early postpartum events (hemorrhage, ICU) | 180 (5.4) |
| **Early postpartum events assessed at 8 weeks** (N=3015) | |
| Child hospitalization or disease | 331 (11.0) |
| Maternal hospitalization or disease | 38 (1.3) |
| Pain or incontinence | 774 (25.7) |
| No sexual activity | 904 (30.0) |
| No breastfeeding | 1411 (46.8) |
| **Late postpartum events assessed at one year** (N=2351) | |
| Child hospitalization or disease | 373 (15.9) |
| Maternal hospitalization or disease | 32 (1.4) |
| New pregnancy | 97 (4.1) |

**Table 6:** Pregnancy, delivery and postpartum events among IGEDEPP cohort respondents Abbreviations: GA, gestation age; NICU, newborn Intensive Care Unit; IUC, Intensive care unit.

|                                               | 8 weeks follow-up (N=3015) | Lost to follow-up during the first 8 weeks (N=295) | p value |
|---|---|---|---|
| Age | 32.7 (4.5) | 32.1 (4.9) | **0.025** |
| Primiparous | 1726 (57.2) | 170 (58) | 0.95 |
| Marital status: Single | 90 (3.0) | 13 (4) | 0.24 |
| Education level: high School or less | 223 (7.4) | 38 (13) | **0.001** |
| Job status: unemployed | 198 (6.6) | 24 (8) | 0.36 |
| Current psychotropic medication | | | |
|     Antidepressant | 48 (1.6) | 5 (2) | 1 |
|     Antiepileptic | 8 (0.3) | 1 (0.3) | 0.57 |
|     Antipsychotic | 5 (0.2) | 0 (0) | 1 |
|     Other | 3 (0.1) | 1 (0.3) | 0.31 |
| Lifetime history of major depressive episode | 1053 (34.9) | 113 (38) | 0.27 |
| Major depressive episode during current pregnancy | 110 (3.7) | 14 (5) | 0.43 |

**Supplementary table 1**: Comparison of patient's characteristics between participants lost to follow-up during the first 8 weeks (N=295) and participants evaluated at W8 (N=2351).

|  | One year follow-up (N=2351) | Lost to follow-up during fist year (N=959) | p value |
|---|---|---|---|
| Age | 32.8 (4.4) | 32.2 (4.7) | **<0.001** |
| Primiparous | 1355 (57.6) | 541 (56.4) | 0.54 |
| Marital status: Single | 64 (2.7) | 39 (4.1) | 0.056 |
| Education level: high School or less | 147 (6.3) | 114 (11.9) | **<0.001** |
| Job status: unemployed | 151 (6.4) | 71 (7.4) | 0.34 |
| Medication: |  |  |  |
|    Antidepressant | 36 (1.5) | 17 (1.8) | 0.73 |
|    Antiepileptic | 6 (0.3) | 3 (0.3) | 0.72 |
|    Antipsychotic | 3 (0.1) | 2 (0.2) | 0.63 |
|    Other | 3 (0.1) | 1 (0.1) | 1 |
| History of depressive episode | 812 (34.5) | 354 (36.9) | 0.21 |
| History of depressive episode during the current pregnancy | 83 (3.5) | 41 (4.3) | 0.36 |

**Supplementary table 2**: comparison between patients lost to follow-up during the first year (N=959) and patients with complete follow-up.

| Centre | B | C | H | J | L | N | P | T | p (Chi-square) |
|---|---|---|---|---|---|---|---|---|---|
| Prevalence of PPD at W8 (%) | 5.9 | 6.8 | 7.0 | 7.3 | 6.7 | 8.6 | 11.5 | 5.2 | 0.049 |

**Supplementary table 3**: Prevalence of PPD at W8 in the different centers participating in the study

| Category | Whole pregnancy (N=3310) N (%) | Early post-partum (N=3015)[a] N (%) | Late post-partum (N=2351)[a] N (%) |
| --- | --- | --- | --- |
| Work | 392 (11.8) | 194 (6.4) | 308 (13.1) |
| Education | 38 (1.2) | 18 (0.6) | 30 (1.3) |
| Wealth | 255 (7.7) | 87 (2.9) | 115 (4.9) |
| Health | 355 (10.7) | 117 (3.9) | 256 (10.9) |
| Bereavement | 202 (6.1) | 83 (2.8) | 179 (7.6) |
| Moving | 161 (4.9) | 71 (2.4) | 95 (4.0) |
| Love relationship | 36 (1.1) | 27 (0.9) | 27 (1.2) |
| Legal issues | 60 (1.8) | 22 (0.73) | 26 (1.1) |
| Social or familial | 443 (13.4) | 248 (8.2) | 199 (8.5) |
| Marital | 207 (6.3) | 153 (5.1) | 249 (10.6) |
| Others | 385 (11.6) | 310 (10.3) | 282 (12.0) |
| **Any** | **1578 (47.7)** | **912 (30.2)** | **1033 (43.9)** |

**Supplementary table 4**: Adult stressful life events during pregnancy and postpartum period among IGEDEPP cohort respondents.
[a]Early post-partum was defined as 8 weeks postpartum, late post-partum was defined as ranging between two months and one year postpartum.